\documentclass[10pt]{article}
\usepackage{stmaryrd}
\usepackage{amsfonts}
\usepackage{bbm}
\usepackage{amscd}
\usepackage{mathrsfs}
\usepackage{latexsym,amssymb,amsmath,amscd,amscd,amsthm,amsxtra,xypic}
\usepackage[dvips]{graphicx}
\usepackage{latexsym,amssymb,amsmath,amscd,amscd,amsthm,amsxtra}

\newtheorem{thm}{Theorem}[section]

\newtheorem{lem}[thm]{Lemma}
\newtheorem{cor}[thm]{Corollary}
\newtheorem{pro}[thm]{Proposition}
\newtheorem{ex}[thm]{Example}
\newtheorem{rmk}[thm]{Remark}
\newtheorem{defi}[thm]{Definition}

\newcommand{\lon }{\,\rightarrow\,}
\newcommand{\be }{\begin{eqnarray*}}
\newcommand{\ee }{\end{eqnarray*}}

\newcommand{\defbe}{\triangleq}
\newcommand{\pf}{\noindent{\bf Proof.}\ }

\newcommand{\huaE}{\mathcal{E}}

\newcommand{\huaJ}{\mathcal{J}}

\newcommand{\huaK}{\mathcal{K}}

\newcommand{\huaT}{\mathcal{T}}

\newcommand{\set}[1]{\left\{#1\right\}}

\newcommand{\pair}[1]{\left\langle #1\right\rangle}
\newcommand{\ppairingE}[1]{\left ( #1\right )_E}

\newcommand{\frkd}{\mathfrak d}

\newcommand{\frkr}{\mathfrak r}

\newcommand{\frkt}{\mathfrak t}

\newcommand{\frkL}{\mathfrak L}

\newcommand{\frkX}{\mathfrak X}

\def\gpd{\,\lower1pt\hbox{$\longrightarrow$}\hskip-.24in\raise2pt
         \hbox{$\longrightarrow$}\,}

\def\qed{\hfill ~\vrule height6pt width6pt depth0pt}

\newcommand{\Lied}{\frkL}

\newcommand{\conpairing}[1]{\left\langle  #1\right\rangle }

\newcommand{\jacobi}[1]{\left\lceil  #1 \right \rceil }
\newcommand{\Dorfman}[1]{\left  \llbracket  #1\right  \rrbracket }

\newcommand{\jet}{\mathfrak{J}}

\newcommand{\jetd}{\mathbbm{d}}
\newcommand{\dev}{\mathfrak{D}}
\newcommand{\pie}{^\prime}
\newcommand{\yi}{\mathbf{1}}
\newcommand{\Id}{\mathrm{Id}}

\newcommand{\dM}{\mathrm{d}}

\newcommand{\Hom}{\mathrm{Hom}}
\newcommand{\gl}{\mathrm{gl}}

\begin{document}
\title{
{Jacobi Quasi-Nijenhuis Algebroids
\thanks
 {
 Research partially supported by NSF of China (10871007), China Postdoctoral Science Foundation, Science Research Foundation for Excellent Young Teachers of College of
Mathematics at Jilin University, US-China CMR Noncommutative
Geometry (10911120391/A0109).
 }
} }
\author{Yunhe Sheng  \\
School of Mathematics, Jilin University,
 Changchun 130012, Jilin, China\\
 and
\\
School of Mathematics, Dalian University of Technology, Dalian
116024, China
\\
email: ysheng888@gmail.com}

\date{}


\maketitle %\tableofcontents

\begin{abstract}
In this paper, for a Jacobi algebroid $(A,\rho)$, by introducing the
notion of Jacobi quasi-Nijenhuis algebroids, which is a
generalization of Poisson quasi-Nijenhuis manifolds introduced in
\cite{Poisson quasi-Nijenhuis} by Sti\'{e}non and Xu, we study
generalized complex structures on the Courant-Jacobi algebroid
$A\oplus A^*$,  which unify generalized complex (contact) structures
on an even(odd)-dimensional manifold.
\end{abstract}

\section{Introduction}
The notion of Poisson quasi-Nijenhuis manifolds was introduced in
\cite{Poisson quasi-Nijenhuis} by Sti\'{e}non and Xu. In
\cite{Poisson quasi-Nijenhuis background}, the author studied
Poisson quasi-Nijenhuis structures with background. One can study
generalized complex structures in term of Poisson quasi-Nijenhuis
structures. Generalized  complex structures were introduced by
Hitchin \cite{Hitchin} and further studied by Gualtieri
\cite{gualtieri} as a bridge of symplectic and complex structures.
Note that only on even-dimensional manifolds, there are generalized
complex structures.  In \cite{Wade contact manifold and g},
Iglesias-Ponte and Wade gave the odd-dimensional analogue of the
concept of generalized complex structures under the name of
generalized contact structures.

Jacobi structures on a manifold $M $ are local Lie algebra
structures \cite{KirillovLocal} on $C^\infty(M)$. It contains a
bi-vector field $\Lambda$ and a vector field $X$ such that
$[\Lambda,\Lambda]=2X\wedge\Lambda$ and $[X,\Lambda]=0$.  In
\cite{generalize Lie bialgebroid}, Iglesias and Marrero introduced
the notion of generalized Lie bialgebroids in such a way that the
base manifold is a Jacobi manifold. The same object was introduced
in \cite{Grabowski marmo1} by Grabowski and Marmo under the name of
Jacobi bialgebroids. Similar as the fact that the double of a Lie
bialgebroid is a Courant algebroid, the double of a generalized Lie
bialgebroid (Jacobi bialgebroid) is a generalized Courant algebroid
(Courant-Jacobi algebroid). These topics are widely studied
 \cite{Contractions}, \cite{Contractions1},
\cite{Jacobi-Nijenhuis algebroid}, \cite{Courant-Nijenhuis},
 \cite{Grabowski marmo1},
\cite{Grabowski marmo2}, \cite{generalize Lie bialgebroid},
  \cite{Nunes da Costa quasi-Jacobi and Jacobi quasi}, \cite{Nunes da Costa
quasi-Jacobi}, \cite{Dirac generalized Lie bi}.

In this paper, for  a Jacobi algebroid $(A,\rho)$, we study Jacobi
quasi-Nijenhuis structures. As an application, we study generalized
complex structures on  the Courant-Jacobi algebroid $A\oplus A^*$,
which unify generalized complex structures on an even-dimensional
manifold and generalized contact structures on an odd-dimensional
manifold. By definition, a  Jacobi quasi-Nijenhuis algebroid is a
quadruple $((A,\rho),\pi,N, \phi)$, where $(A,\rho)$ is a Jacobi
algebroid, $\pi\in\Gamma(\wedge^2A)$ is a Jacobi bi-vector field,
$N\in\Gamma(A^*\otimes A)$ is compatible with $\pi$, and
$\phi\in\Gamma(\wedge^3A^*)$ satisfying $\jetd\phi=0$ and
$\jetd(i_N\phi)=0$, such that the Nijenhuis torsion $T(N)$ of $N$
can be expressed as
$$
T(N)(X,Y)=\pi^\sharp(i_{X\wedge Y}\phi),\quad\forall
~X,Y\in\Gamma(A).
$$
We generalize some  well known results and formulas which hold in
the case of Poisson quasi-Nijenhuis manifolds. The biggest
obstruction is that in the frame work of ``Jacobi" world, the
differential and the Lie derivative are no longer derivations with
respect to the wedge product, $\wedge$. A Generalized complex
structure  is defined in the usual way, it is a bundle map
$\huaJ:A\oplus A^*\longrightarrow A\oplus A^*$ preserving the
canonical pairing and satisfying $\huaJ^2=-\Id$ as well as the
integrability condition, which is expressed in term of the
Courant-Jacobi bracket. Since the usual Courant algebroid and
$\huaE^1(M)$ are special Courant-Jacobi algebroids, thus it unifies
generalized complex structures on an even-dimensional manifold and
generalized contact structures on an odd-dimensional manifold.

The paper is organized as follows. In Section 2 we proved that there
is a one-to-one correspondence between quasi-Jacobi bialgebroids and
quasi-Manin triples. In Section 3 we introduce the notion of Jacobi
quasi-Nijenhuis algebroids and give the relation with quasi-Jacobi
bialgebroids. In Section 4 we study generalized complex structures
on the Courant-Jacobi algebroid $A\oplus A^*$. We prove that there
is also a one-to-one correspondence between generalized complex
structures and Jacobi quasi-Nijenhuis algebroids satisfying a
homomorphism condition. In Section 5 we study generalized complex
structures on $TM\oplus T^*M$ for an even-dimensional manifold $M$
and we will see that how a conformal symplectic structure is
involved in a generalized complex structure. In Section 6 we study
generalized complex structures on $\huaE^1(M)$ for an
odd-dimensional manifold $M$. Since $(TM\oplus\mathbb R,\Id)$ is a
natural Jacobi algebroid, we recover the notion of generalized
contact structures introduced in \cite{Wade contact manifold and g}.
Some examples are also discussed.

{\bf Notations:} We denote the usual Lie bracket of vector fields
or the Lie bracket on a Lie algebroid by $[\cdot,\cdot]$, the
bracket of the Schouten-Jacobi algebra decided by a Jacobi
algebroid by $\Dorfman{\cdot,\cdot}$, the bracket on a
Courant-Jacobi algebroid by $\jacobi{\cdot,\cdot}$. $\dM$ is the
usual deRham differential or the differential associated with a
Lie algebroid. $\jetd$ is the differential associated with a
Jacobi algebroid. For any $X\in\Gamma(A)$, where $(A,\rho)$ is a
Jacobi algebroid, $L_X$ is the usual Lie derivative decided by the
Lie algebroid structure and $\Lied_X$ is the Lie derivative
decided by the Jacobi algebroid structure. $\yi$ is the constant
function with the value $1$. $\Id$ is the identity map if there is
no special explanation.

{\bf Acknowledgements:} We would like to give our special thanks to
Zhangju Liu for very helpful comments and also give our warmest
thanks to Chenchang Zhu for the help during we stayed in Courant
Research Center, G$\ddot{\rm{o}}$ttingen, where a part of work was
done. We also give our warmest thanks to the referees for many
helpful suggestions and pointing out typos and erroneous statements.

\section{Quasi-Manin triples}
%Throughout this paper, $M$ is a smooth manifold, $\dM$ is the usual differential operator on forms, $E\longrightarrow M$ is  a vector bundle.  In this section,
%we introduce some notions of generalized complex structures introduced by Gualtieri in \cite{gualtieri}, generalized contact structures introduced by
%Iglesias-Ponte and Wade in \cite{Wade contact manifold and g} and $E$-Courant algebroids introduced by the authors in \cite{CLS2}.
A Lie algebroid over a manifold $M$ is a vector bundle
$A\longrightarrow M$ together with a Lie bracket $[\cdot,\cdot]$
on the section space $\Gamma(A)$ and a bundle map
$a:A\longrightarrow TM$, called the anchor, satisfying the
compatible condition:
$$
[X,fY]=f[X,Y]+a(X)(f)Y,\quad\forall~X,~Y\in\Gamma(A),~f\in
C^\infty(M).
$$
We usually denote a Lie algebroid by $(A,[\cdot,\cdot],a)$, or $A$
if there is no confusion.
%The coboundary operator $\dM:\Gamma(\wedge^{\bullet} A^*)\longrightarrow\Gamma(\wedge^{\bullet+1} A^*)$ is given by
%\begin{eqnarray*} \dM c (X_0,\cdots,X_k)&=&\sum_{i=0}^k(-1)^ia(X_i)c(X_0,\cdots,\widehat{X_i},\cdots,X_k)\\
%&&+\sum_{i<j}(-1)^{i+j}c([X_i,X_\huaJ],X_0,\cdots,\widehat{X_i},\cdots,\widehat{X_\huaJ},\cdots,X_k), \end{eqnarray*}
%for any $c\in\Gamma(\wedge^{k} A^*)$ and $X_0,\cdots,X_k\in\Gamma(A)$. For any $X\in\Gamma(A)$, the Lie derivative
%$L_X:\Gamma(\wedge^kA^*)\longrightarrow\Gamma(\wedge^kA^*)$ is defined by the Cartan formula:
%\begin{equation} L_Xc=i_X\dM c+\dM i_X c. \end{equation}
For a $(1,1)$-tensor $N\in\Gamma(A^*\otimes A)$, the Nijenhuis
torsion $T(N):\wedge^2A\longrightarrow A$ is defined by
         \begin{equation}\label{eqn:Nijenhuis torsion}
            T(N)(X,Y)=[NX,NY]-N([NX,Y]+[X,NY]-N[X,Y]),\quad\forall~X,Y\in\Gamma(A).
        \end{equation}
        If $T(N)=0$, $N$ is called a Nijenhuis operator on the Lie
        algebroid $A$.
We can also introduce a new bracket $[\cdot,\cdot]_N$  on
$\Gamma(A)$ which is  defined as follows:
\begin{equation}\label{eqn:Nijenhuis bracket}
            [X,Y]_N=[NX,Y]+[X,NY]-N[X,Y],\quad\forall ~X, Y
            \in\Gamma(A).
\end{equation}
If $N$ is  a Nijenhuis operator, $[\cdot,\cdot]_N$ is also a Lie
bracket and $N$ is a Lie algebroid morphism from Lie algebroid
$(A,[\cdot,\cdot]_N,a\circ N)$ to Lie algebroid
$(A,[\cdot,\cdot],a)$.

For any $\pi\in\Gamma(\wedge^2A)$ and
$\sigma\in\Gamma(\wedge^2A^*)$, $\pi^\sharp:A^*\longrightarrow A$
and $\sigma_\flat:A\longrightarrow A^*$ are given by
$$
\pi^\sharp(\xi)(\eta)=\pi(\xi,\eta),\quad\sigma_\flat(X)(Y)=\sigma(X,Y),\quad\forall~\xi,\eta\in\Gamma(A^*),\forall~X,Y\in\Gamma(A).
$$
For any $N\in\Gamma(A^*\otimes A)$ and $\pi\in\Gamma(\wedge^2A)$,
$\pi_N\in\Gamma(\wedge^2A)$ is defined by
$$
\pi_N(\xi,\eta)=\eta(N\pi^\sharp(\xi)),\quad\forall~\xi,\eta\in\Gamma(A^*).
$$
%It is skew-symmetric if and only if $N\circ\pi^\sharp=\pi^\sharp\circ N^*$. \vspace{3mm}

A Jacobi algebroid is a Lie algebroid $(A,[\cdot,\cdot],a)$
together with a 1-cocycle $\phi_0\in\Gamma(A^*)$ and we denote it
by $(A,\phi_0)$. There is a $\phi_0$-bracket
$[\cdot,\cdot]_{\phi_0}$ on $\Gamma(\wedge^\bullet A)$, which is
given by
\begin{equation}
[P,Q]_{\phi_0}=[P,Q]+(-1)^{p+1}(p-1)P\wedge
i_{\phi_0}Q-(q-1)i_{\phi_0}P\wedge Q,
\end{equation}
for any $P\in \Gamma(\wedge^p A)$ and $Q\in\Gamma(\wedge^q A)$.
%It satisfies \begin{equation}  [P,Q\wedge R]_{\phi_0}=[P,Q]_{\phi_0}\wedge R+(-1)^{q(p+1)}Q\wedge[P,R]_{\phi_0}-(i_{\phi_0}P)\wedge Q\wedge R. \end{equation}
The $\phi_0$-differential $\dM^{\phi_0}$ and the $\phi_0$-Lie
derivative $L^{\phi_0}_X$ \cite{generalize Lie bialgebroid} are
defined by
$$
\dM^{\phi_0} c=\dM c+\phi_0\wedge c,\quad
L^{\phi_0}_X=i_X\dM^{\phi_0} +\dM^{\phi_0}i_X.
$$
In fact, a Jacobi algebroid $(A,\phi_0)$ is equivalent to the Lie
algebroid $(A,[\cdot,\cdot],a)$ together with a representation
$\rho:A\longrightarrow\dev(M\times \mathbb R)$ on the trivial line
bundle $M\times \mathbb R$, where $\dev(M\times \mathbb R)$ is the
gauge Lie algebroid of  $M\times \mathbb R$. The representation is
given by
\begin{equation}\label{eqn:rep}
\rho(u)(f)=a(u)f+\phi_0(u)f,\quad\forall~u\in\Gamma(A),~f\in
C^\infty(M)=\Gamma(M\times \mathbb R).
\end{equation}
One can easily prove that $\rho$ is a representation if and only if
$\phi_0$ is a 1-cocycle. More generally we have
\begin{lem}
For any $\theta\in\Gamma(A^*\otimes(M\times\gl(n)))$, i.e.
$\gl(n)$-valued $1$-form on $A$, $\rho=a+\theta$ is a representation
on $M\times \mathbb R^n$ if and only if $\theta$ satisfies the
Maurer-Cartan equation, more precisely,
$$
 \dM\theta+\frac{1}{2}[\theta\wedge\theta]=0.
$$
\end{lem}
\pf By straightforward computations, we have
\begin{eqnarray*}
[\rho(X),\rho(Y)]&=&[a(X)+\theta(X),a(Y)+\theta(Y)]\\
&=&[a(X),a(Y)]+[\theta(X),\theta(Y)]+a(X)\theta(Y)-a(Y)\theta(X)
\end{eqnarray*}
On the other hand, $\rho([X,Y])=a([X,Y])+\theta([X,Y]), $ therefore,
after comparing the values in $TM$ and $ M\times \gl(n)$, we obtain
the required result. \qed

Conversely, for the Lie algebroid $(A,[\cdot,\cdot],a)$ and a
representation $\rho:A\longrightarrow\dev(M\times\mathbb R)$, denote
by $\jetd:\Gamma(\wedge^\bullet
A^*)\longrightarrow\Gamma(\wedge^{\bullet+1}A^*)$ the associated
differential operator, i.e.
\begin{eqnarray}\label{eqn:jerd}
\nonumber\jetd c
(X_0,\cdots,X_k)&=&\sum_{i=0}^k(-1)^i\rho(X_i)c(X_0,\cdots,\widehat{X_i},\cdots,X_k)\\
&&+\sum_{i<j}(-1)^{i+j}c([X_i,X_j],X_0,\cdots,\widehat{X_i},\cdots,\widehat{X_j},\cdots,X_k).
\end{eqnarray}
Then we can obtain a 1-cocycle $\jetd\yi\in\Gamma(A^*)$. Obviously,
if
 the representation $\rho$ is given by $(\ref{eqn:rep})$, then
 $$
\phi_0=\jetd\yi,\quad
\jetd\omega=\dM\omega+\phi_0\wedge\omega,\quad\forall~
\omega\in\Gamma(\wedge^kA^*).
 $$
Therefore, we have $\jetd=\dM^{\phi_0}$,  the $\phi_0$-differential.
Consequently for any $X\in\Gamma(A)$, we can define the Lie
derivative
$\Lied_X:\Gamma(\wedge^kA^*)\longrightarrow\Gamma(\wedge^kA^*)$ by
Cantan formula:
$$
\Lied_X=i_X\circ\jetd+\jetd\circ i_X.
$$
Obviously, we have $\Lied_X\omega=L_X\omega+\phi_0(X)\omega$, which
implies $\Lied_X=L^{\phi_0}_X$,  the
 $\phi_0$-Lie derivative.

\begin{rmk} We should be very careful that since $\jetd$ is no longer a
derivation, $\Lied_X$ is not a derivation. Therefore, the induced
Lie derivative
$\Lied_X:\Gamma(\wedge^kA)\longrightarrow\Gamma(\wedge^kA)$ is also
not a derivation. This Lie derivative is exactly the foundation of
the $\phi_0$-bracket introduced in \cite{generalize Lie
bialgebroid}. Certainly, by this Lie derivative we can only define
the $\phi_0$-bracket of a $1$-vector field and a $k$-vector field,
and then by some rules one can obtain the bracket of any $l$-vector
field and any $k$-vector field,  see also \cite{Grabowski marmo1}
and \cite{Grabowski marmo2} for more details.
\end{rmk}

{\bf Convention:} We denote a Jacobi algebroid by $(A,\rho)$ and the
associated Schouten-Jacobi algebra by $(\Gamma(\wedge^\bullet
A),\Dorfman{\cdot,\cdot})$.

The notion of Courant-Jacobi algebroids was introduced in
\cite{Grabowski marmo2}. In \cite{generalize Lie bialgebroid}, the
authors proved that they are the same as  generalized Courant
algebroids. They are generalizations of Courant algebroids
introduced in \cite{LWXmani}, see also \cite{Roytenberg}. In fact,
Courant algebroids and Courant-Jacobi algebroids are all special
cases of $E$-Courant algebroids introduced in \cite{CLS2}, where $E$
is a vector bundle.

\begin{defi}\label{defi:CourantJacobi}
A Courant-Jacobi algebroid is  a vector bundle $\huaK$ over $ M$
together with
\begin{itemize}
\item[\rm(1)] a nondegenerate symmetric bilinear form
$\pair{\cdot,\cdot}$ on the bundle; \item[\rm(2)] a bilinear
operator $\circ$ on $\Gamma(\huaK)$ such that
$(\Gamma(\huaK),\circ)$ is a Leibniz algebra; \item[\rm(3)] a bundle
map $\kappa:\huaK\lon TM\times \mathbb R$ which is a homomorphism
into the Lie algebroid of first order differential operators
satisfying the following properties,
$$\rm(a).~ \langle Y\circ X,X\rangle=\langle Y,X\circ X\rangle, \qquad\rm(b).~\kappa(X)\langle Y,Y\rangle=2\langle X\circ
Y,Y\rangle.$$
\end{itemize}
\end{defi}

\begin{defi}
A quasi-Jacobi bialgebroid is a triple $((A,\rho),\delta,\phi)$
consisting of a Jacobi algebroid $(A,\rho)$, a degree $1$
derivation $\delta$ of the Schouten-Jacobi algebra
$(\Gamma(\wedge^\bullet A),\Dorfman{\cdot,\cdot})$ and an element
$\phi\in\Gamma(\wedge^3 A)$ such that
$\delta^2=\Dorfman{\phi,\cdot}$ and $\delta\phi=0$.
\end{defi}
\begin{defi}
A quasi-Manin triple is a triple $(\huaK,A,B)$, where $\huaK$ is a
Courant-Jacobi algebroid, $A\subset\huaK$ is a Dirac  structure and
$B$ is its transversal isotropic complement.
\end{defi}
\begin{rmk}
In \cite{Nunes da Costa quasi-Jacobi}, the notion of quasi-Jacobi
bialgebroids has already been introduced, which is motivated by
\cite{Roytenberg quasi}. Our definition is motivated by \cite{Xu
quasi-Lie bi}. One can easily recover the six conditions in their
definition and some of the constructions are given in the proof of
the next theorem.
\end{rmk}
\begin{thm}\label{thm:quasi Liebi}
There is a one-to-one correspondence between  quasi-Jacobi
bialgebroids and quasi-Manin triples.
\end{thm}
\pf Let $((A,\rho),\delta,\phi)$ be a quasi-Jacobi bialgebroid.
Define the bundle map $\rho_*:A^*\longrightarrow TM\oplus \mathbb
R$  by
\begin{equation}\label{defi:rho8}
\rho_*(\xi)(f)=\xi(\delta(f)),\quad\forall~\xi\in\Gamma(A^*),f\in
C^\infty(M).
\end{equation}
Introduce a bracket $[\cdot,\cdot]_*$ on $\Gamma(A^*)$ by
$$
[\xi,\eta]_*(X)=\rho_*(\xi)(\eta(X))-\rho_*(\eta)(\xi(X))-\delta(X)(\xi,\eta).
$$
$\rho_*$ is not a homomorphism but we have
$$
\rho_*[\xi,\eta]_*=[\rho_*(\xi),\rho_*(\eta)]-\rho(\phi(\xi,\eta)).
$$
Therefore, in general, $(A^*,[\cdot,\cdot]_*,\rho_*)$ is not a
Jacobi algebroid. Let $\kappa:A\oplus A^*\longrightarrow
TM\oplus\mathbb R$ be the bundle map given by
$$
\kappa(X+\xi)=\rho(X)+\rho_*(\xi).
$$
Define a bracket $ \jacobi{\cdot,\cdot}$ on $\Gamma(A\oplus A^*)$ by
\begin{eqnarray*}
\jacobi{X,Y}&=&[X,Y],\qquad\forall~X,Y\in\Gamma(A),\\
\jacobi{\xi,\eta}&=&[\xi,\eta]_*+\phi(\xi,\eta,\cdot),\quad\forall~~\xi,~\eta\in\Gamma(A^*),\\
\jacobi{X,\xi}&=&i_X\jetd\xi-i_\xi\delta(X)+\jetd(\xi(X)),\\
\jacobi{\xi,X}&=&-i_X\jetd\xi+i_\xi\delta(X)+\delta(\xi(X)).
\end{eqnarray*}
%where $\jetd_*:\Gamma(\wedge^kA)\longrightarrow\Gamma(\wedge^{k+1}A)$ is  decided by $[\cdot,\cdot]_*$ and $\rho_*$ and given by (\ref{eqn:jerd}).
Then $(A\oplus
A^*,\pair{\cdot,\cdot},\jacobi{\cdot,\cdot},\kappa)$ is a
Courant-Jacobi algebroid such that $A$ is a Dirac structure and
$A^*$ is its transversal isotropic complement.

Conversely, assume that
$(\huaK,\pair{\cdot,\cdot},\jacobi{\cdot,\cdot},\kappa)$ is a
Courant-Jacobi algebroid and $A $ is a Dirac structure with a
transversal isotropic complement $B$, by using the pairing, we can
identify $B$ with $A^*$. Let $\rho=\kappa|_A$ be the restriction of
$\kappa$ on $A$, then $(A,\rho)$ is a Jacobi algebroid. $\phi\in
\Gamma(\wedge^3A)$ is defined by
\begin{equation}
\phi(\xi,\eta,\gamma)=2\pair{\jacobi{\xi,\eta},\gamma},\quad\forall~\xi,\eta,\gamma\in\Gamma(B).
\end{equation}
Let $\rho_B=\kappa|_B$ be the restriction of $\kappa$ to $B$ and
$[\cdot,\cdot]_B$ be the bracket on $\Gamma(B) $ such that
\begin{equation}\label{eqn:bracket quasi}
\jacobi{\xi,\eta}-[\xi,\eta]_B\in\Gamma(A).
\end{equation}
Define $\delta:\Gamma(\wedge^\bullet
A)\longrightarrow\Gamma(\wedge^{\bullet+1}A)$ by $\rho_B$ and the
bracket $[\cdot,\cdot]_B$ as (\ref{eqn:jerd}), then
$((A,\rho),\delta,\phi)$ is a quasi-Jacobi bialgebroid. \qed

\section{Jacobi quasi-Nijenhuis algebroids}
 A {\bf Jacobi bi-vector field} on a Jacobi algebroid $(A,\rho)$ is a bi-vector field $\pi\in\Gamma(\wedge^2A)$
 satisfying $$ \Dorfman{\pi,\pi}=0.$$
\begin{rmk}\label{rmk:Jacobi bi-vector field}
It is called a Jacobi bi-vector field because in the case that
$A=TM\times\mathbb R$ and the Lie algebroid structure on
$TM\times\mathbb R$ is given by
\begin{equation}\label{eqn:bracket TMR}
[X+f,Y+g]=[X,Y]+Xg-Yf,\quad\forall~X+f,Y+g\in\frkX(M)\oplus
C^\infty(M),
\end{equation}
a bi-vector field is a pair $(\Lambda,X)$, where
$\Lambda\in\frkX^2(M)$ and $X\in\frkX(M)$, and $(\Lambda,X)$ is a
Jacobi bi-vector field if and only if it is a Jacobi structure on
$M$. See \cite{Triangular generalized} and \cite{Jacobi-Nijenhuis
algebroid} for more details.
\end{rmk}

On $\Gamma(A^*)$, we can introduce a Lie bracket
$\Dorfman{\cdot,\cdot}_\pi$, which is induced by a Jacobi bi-vector
field $\pi$:
\begin{equation}\label{eqn:bracket pi}
\Dorfman{\xi,\eta}_\pi=-\jetd(\pi(\xi,\eta))+\Lied_{\pi^\sharp(\xi)}\eta-\Lied_{\pi^\sharp(\eta)}\xi,\quad
\forall~\xi,\eta\in\Gamma(A^*).
\end{equation}

\begin{pro}\label{lem:A8pi}
Let $(A,\rho)$ be a Jacobi algebroid, $\pi\in\Gamma(\wedge^2A)$ is
a Jacobi bi-vector field if and only if
$(A^*,\rho\circ\pi^\sharp)$ is a Jacobi algebroid, where the Lie
algebroid structure on $A^*$ is given by
$(A^*,\Dorfman{\cdot,\cdot}_\pi,a\circ\pi^\sharp)$. In this case,
we have $\jetd_*\yi=-\pi^\sharp(\jetd\yi)$.
\end{pro}
\pf Since we also have the well known formula:
\begin{equation}\label{eqn:main}
\pi^\sharp\Dorfman{\xi,\eta}_\pi-[\pi^\sharp(\xi),\pi^\sharp(\eta)]=\frac{1}{2}\Dorfman{\pi,\pi}(\xi,\eta),
\end{equation}
it follows that $\Dorfman{\cdot,\cdot}_\pi$ is a Lie bracket if
and only if $\pi$ is a Jacobi bi-vector field. In this case, it is
obvious that $\rho\circ\pi^\sharp $ is a representation of the Lie
algebroid $(A^*,\Dorfman{\cdot,\cdot}_\pi,a\circ\pi^\sharp)$. For
any $\xi\in\Gamma(A)$, we have
$$
\xi(\jetd_*\yi)=\rho\circ\pi^\sharp(\xi)(\yi)=\jetd\yi(\pi^\sharp(\xi))=-\xi(\pi^\sharp\jetd\yi),
$$
which implies $\jetd_*\yi=-\pi^\sharp(\jetd\yi)$ and the proof is
finished.\qed

\begin{defi}
Let $(A,\rho)$ be a Jacobi algebroid, a Jacobi bi-vector field $\pi$
and a $(1,1)$-tensor $N:A\longrightarrow A$ are compatible if the
following two conditions are satisfied: $$
                    N\circ\pi^\sharp=\pi^\sharp\circ N^*\quad and
                    \quad C(\pi,N)= 0,
      $$
 where
                \begin{equation}
                   C(\pi,N)(\xi,\eta)\triangleq \Dorfman{\xi,\eta}_{\pi_N}-(\Dorfman{N^* \xi,\eta}_\pi+ \Dorfman{\xi,N^*\eta}_\pi -
                   N^*\Dorfman{\xi,\eta}_\pi),\quad \forall
                   ~\xi,\eta\in\Gamma(A^*).
                   \end{equation}
In the case where $N$ is a Nijenhuis operator, i.e. $T(N)=0$, the
triple  $((A,\rho),\pi,N)$ is said to be a {\bf Jacobi-Nijenhuis
algebroid}.
\end{defi}
\begin{rmk}
The notion of a Jacobi-Nijenhuis algebroid has already appeared in
\cite{Jacobi-Nijenhuis algebroid}, where the author use the
condition $\Dorfman{\pi,\pi_N}=0$ instead of $C(\pi,N)= 0$. In
fact, if $C(\pi,N)= 0$, we can deduce that
$\Dorfman{\pi,\pi_N}=0$, this is given by the next lemma
\end{rmk}

\begin{lem}
Let $(A,\rho)$ be a Jacobi algebroid, the Jacobi bi-vector field
$\pi$ is compatible with the $(1,1)$-tensor $N$, then we have
$$
\Dorfman{\pi,\pi_N}=0.
$$
\end{lem}
\pf By (\ref{eqn:main}), we can obtain
$$
\Dorfman{\pi,\pi_N}(\xi,\eta)=\pi^\sharp\Dorfman{\xi,\eta}_{\pi_N}+\pi^\sharp\circ
N^*\Dorfman{\xi,\eta}_\pi
-[\pi^\sharp(\xi),\pi^\sharp(N^*\eta)]-[\pi^\sharp(N^*\xi),\pi^\sharp(\eta)].
$$
If $\pi$ and $N$ are compatible, we have
$$
\Dorfman{\xi,\eta}_{\pi_N}=\Dorfman{N^* \xi,\eta}_\pi+
\Dorfman{\xi,N^*\eta}_\pi - N^*\Dorfman{\xi,\eta}_\pi.
$$
The conclusion follows from the fact that
$\pi^\sharp\Dorfman{\xi,\eta}_{\pi}=[\pi^\sharp(\xi),\pi^\sharp(\eta)]$.\qed
\vspace{3mm}

The degree $0$ derivation $i_N$ of $\Gamma(\wedge^\bullet A^*)$ is
defined by
$$
(i_N\omega)(X_1,\cdots,X_k)=\sum_{i=1}^k\omega(X_1,\cdots,NX_i,\cdots,X_k),\quad\forall~
\omega\in\Gamma(\wedge^kA^*),
$$
and we obtain a degree $1$ differential operator
$\jetd_N:\Gamma(\wedge^\bullet
A^*)\longrightarrow\Gamma(\wedge^{\bullet+1} A^*)$ by the
following formula:
$$
\jetd_N=i_N\circ\jetd-\jetd\circ i_N.
$$

\begin{defi}
A {\bf Jacobi quasi-Nijenhuis algebroid} is a quadruple
$((A,\rho),\pi,N, \phi)$, where $(A,\rho)$ is a Jacobi algebroid,
$\pi\in\Gamma(\wedge^2A)$ is a Jacobi bi-vector field,
$N\in\Gamma(A^*\otimes A)$ is compatible with $\pi$, and
$\phi\in\Gamma(\wedge^3A^*)$ satisfying $\jetd\phi=0$ and
$\jetd(i_N\phi)=0$, such that
\begin{equation}\label{tn}
T(N)(X,Y)=\pi^\sharp(i_{X\wedge Y}\phi),\quad\forall
~X,Y\in\Gamma(A).
\end{equation}
\end{defi}

\begin{thm}\label{thm:quasi-Jacobibi and Jacobi quasi N}
The quadruple $((A,\rho),\pi,N, \phi)$ is a Jacobi quasi-Nijenhuis
algebroid if and only if
$((A^*,\rho\circ\pi^\sharp),\jetd_N,\phi)$ is a quasi-Jacobi
bialgebroid and $\jetd\phi=0$, where the Lie algebroid structure
on $A^*$ is given by
$(A^*,\Dorfman{\cdot,\cdot}_\pi,a\circ\pi^\sharp)$.
\end{thm}
We need the following two lemmas to prove the theorem.

\begin{lem}\label{lem:pi and N}
Let $(A,\rho)$ be a Jacobi algebroid. For a Jacobi bi-vector field
$\pi$ and a $(1,1)$-tensor $N:A\longrightarrow A$, the
differential operator $\jetd_N$ is a derivation of the bracket
$\Dorfman{\cdot,\cdot}_\pi$ if and only if $\pi$ and $N$ are
compatible.
\end{lem}
\pf  This lemma is a generalization of Proposition 3.2 in
\cite{PN2}, where one only need to prove that it holds for
functions and $1$-forms since it is a derivation with respect to
the wedge product, $\wedge$. Here one can prove similarly that
$\jetd_N$ is a derivation for functions and $1$-forms, but since
$\jetd_N$ is no longer a derivation with respect to the wedge
product, $\wedge$, we can not say that it holds in general
directly. But we will see that the obstruction of $\jetd_N$ to be
a derivation is controlled by lower degree elements, therefore, we
can still obtain that $\jetd_N$ is a derivation. In fact, since
$\jetd\omega=\dM \omega+\jetd\yi\wedge\omega$, for any
$P\in\Gamma(\wedge^pA^*)$, we have
\begin{equation}\label{eqn:dnp}
\jetd_NP=\dM_NP+(i_N\jetd\yi)\wedge P.
\end{equation}
Thus, for any $ Q\in\Gamma(\wedge^qA^*)$, we have
$$
\jetd_N(P\wedge Q)=(\jetd_NP)\wedge
Q+(-1)^pP\wedge\jetd_NQ-(i_N\jetd\yi)\wedge P\wedge Q.
$$
On the other hand, by Proposition \ref{lem:A8pi}, for any $
R\in\Gamma(\wedge^rA^*)$, we have
$$
\Dorfman{P,Q\wedge R}_\pi=\Dorfman{P,Q}_\pi\wedge
R+(-1)^{q(p+1)}Q\wedge\Dorfman{P,R}_\pi-(i_{-\pi^\sharp(\jetd\yi)}P)\wedge
Q\wedge R.
$$
Therefore, by direct computation, we have
\begin{eqnarray*}
&&\jetd_N\Dorfman{P,Q\wedge R}_\pi-\Dorfman{\jetd_NP,Q}_\pi\wedge
R+(-1)^p\Dorfman{P,\jetd_N(Q\wedge R)}_\pi\\
&=&\big(\jetd_N\Dorfman{P,Q}_\pi-\Dorfman{\jetd_NP,Q}_\pi+(-1)^p\Dorfman{P,\jetd_NQ}_\pi\big)\wedge
R\\
&&+(-1)^{p(q+1)}Q\wedge\big(\jetd_N\Dorfman{P,
R}_\pi-\Dorfman{\jetd_NP,R}_\pi+(-1)^p\Dorfman{P,\jetd_N
R}_\pi\big)\\
&&-\big(\jetd_N\Dorfman{P,\yi}_\pi-\Dorfman{\jetd_NP,\yi}_\pi+(-1)^p\Dorfman{P,\jetd_N\yi}_\pi\big)\wedge
Q\wedge R.
\end{eqnarray*}
This completes the proof. \qed

\begin{lem}\label{lem:dN}
Let $(A,\rho)$ be a Jacobi algebroid. A Jacobi bi-vector field
$\pi$ and a $(1,1)$-tensor $N$ are compatible. Then
$\jetd_N^2=\Dorfman{\phi,\cdot}_\pi$ is equivalent  to (\ref{tn})
and $\pi^\sharp\circ(\jetd\phi)_\flat=0,$ where
$(\jetd\phi)_\flat:\wedge^3A\longrightarrow A^*$ is the bundle map
defined by $ (\jetd\phi)_\flat(X,Y,Z)=i_{X\wedge Y\wedge
Z}\jetd\phi. $
\end{lem}
\pf By similar computations as in \cite{Poisson quasi-Nijenhuis},
we can easily obtain $\jetd_N^2-\Dorfman{\phi,\cdot}_\pi$ vanishes
on $0$- and exact $1$-forms if and only if $
T(N)(X,Y)=\pi^\sharp(i_{X\wedge Y}\phi) $ and
$\pi^\sharp\circ(\jetd\phi)_\flat=0$. But we should be very
careful that  $\jetd_N^2$ and $ \Dorfman{\phi,\cdot}_\pi$ are no
longer
 derivations with respect to the wedge product, $\wedge$, next we prove that we can
 still get $\jetd_N^2=\Dorfman{\phi,\cdot}_\pi$.
By (\ref{eqn:dnp}), for any $P\in\Gamma(\wedge^pA^*),
Q\in\Gamma(\wedge^qA^*)$, we have
$$
\jetd_N^2(P\wedge Q)=(\jetd_N^2P)\wedge
Q+P\wedge(\jetd_N^2Q)-(\dM_Ni_N\jetd\yi)\wedge P \wedge Q.
$$
On the other hand, we have
$$
\Dorfman{\phi,P\wedge Q}=\Dorfman{\phi,P}\wedge
Q+P\wedge\Dorfman{\phi,Q}-(i_{-\pi^\sharp(\jetd\yi)}\phi)\wedge
P\wedge Q.
$$
We only need to show
$$i_{\pi^\sharp(\jetd\yi)}\phi=-\dM_Ni_N\jetd\yi.$$
By direct computation, for any $X,Y\in\Gamma(A)$, we have
\begin{eqnarray*}
i_{\pi^\sharp(\jetd\yi)}\phi(X,Y)&=&\phi(X,Y)(\pi^\sharp(\jetd\yi))=-\jetd\yi(\pi^\sharp(\phi(X,Y))),\\
\dM_Ni_N\jetd\yi(X,Y)&=&NX\jetd\yi(NY)-NY\jetd\yi(NX)-\jetd\yi(N[X,Y]_N)\\
&=&\jetd\yi([NX,NY]-N[X,Y]_N)=\jetd\yi(\pi^\sharp(\phi(X,Y))).
\end{eqnarray*}
This completes the proof.\qed

{\bf The proof of Theorem \ref{thm:quasi-Jacobibi and Jacobi quasi
N}:}  By Proposition \ref{lem:A8pi}, $\pi$ is a Jacobi bi-vector
field is equivalent to that $(A^*,\rho\circ\pi^\sharp)$ is a
Jacobi algebroid. By   Lemma \ref{lem:pi and N}, $\jetd_N$ is a
derivation is equivalent to $\pi$ and $N$ are compatible. If
$\jetd(i_N\phi)=0$ and $\jetd\phi=0$, we have
$\jetd_N\phi=i_N\jetd\phi-\jetd i_N\phi=0$. Conversely, if
$\jetd_N\phi=\jetd\phi=0$, we have $\jetd i_N\phi=0$. By Lemma
\ref{lem:dN}, the proof is finished. \qed
\begin{thm}
Let $((A,\rho),\pi,N, \phi)$ be a Jacobi quasi-Nijenhuis
algebroid, then we have
$$
 \Dorfman{\pi_N,\pi_N}(\xi,\eta)=-2\pi^\sharp(i_{\pi^\sharp(\xi)\wedge\pi^\sharp(\eta)}\phi).
$$
\end{thm}
\pf By (\ref{eqn:main}), for any $\xi,\eta\in\Gamma(A^*)$, we have
\begin{eqnarray*}
\frac{1}{2}\Dorfman{\pi_N,\pi_N}(\xi,\eta)&=&N\circ\pi^\sharp
\big(\Dorfman{N^*\xi,\eta}_\pi+\Dorfman{\xi,N^*\eta}
-(\Lied_{\pi^\sharp(\xi)}N^*\eta-\Lied_{\pi^\sharp(\eta)}N^*\xi-\jetd\pi(N^*\xi,\eta))\big)\\
&&-\pi^\sharp[N^*\xi,N^*\eta]_\pi\\
&=&N\circ\pi^\sharp\big(\Dorfman{N^*\xi,\eta}_\pi+\Dorfman{\xi,N^*\eta}_\pi-N^*\Dorfman{\xi,\eta}_\pi\big)-\pi^\sharp[N^*\xi,N^*\eta]_\pi\\
&=&-T(N)(\pi^\sharp(\xi),\pi^\sharp(\eta))\\
&=&-\pi^\sharp(i_{\pi^\sharp(\xi)\wedge\pi^\sharp(\eta)}\phi).
\end{eqnarray*}
The second equality holds is because $C(\pi,N)=0$. Since $\pi$ is a
Jacobi bi-vector field, we get the third equality. The last equality
follows from the definition of  a Jacobi quasi-Nijenhuis algebroid.
\qed

\section{Generalized complex structures}

Let $(A,\rho)$ be a Jacobi algebroid. There is a natural pairing
$\pair{\cdot,\cdot}$ on $A\oplus A^*$ which is given by
\begin{equation}\label{eqn:pair of Jacobi bi}
\pair{X+\xi,Y+\eta}=\frac{1}{2}\big(\xi(Y)+\eta(X)\big),\quad\forall
~X,Y\in\Gamma(A),\xi,\eta\in\Gamma(A^*).
\end{equation}
 and we
can introduce a bracket on the section space
$\Gamma(A)\oplus\Gamma(A^*)$ which is given by
\begin{equation}\label{eqn:bracket theta 1}
\jacobi{X+\xi,Y+\eta}=[X,Y]+\Lied_X\eta-\Lied_Y\xi+\jetd(\xi(Y)).
\end{equation}
Obviously, $(A\oplus
A^*,\pair{\cdot,\cdot},\jacobi{\cdot,\cdot},\rho)$ is a
Courant-Jacobi algebroid, where $\rho(X+\xi)=\rho(X)$. In this
section we study generalized complex structures on this
Courant-Jacobi algebroid and we will see that they are related with
Jacobi quasi-Nijenhuis algebroids in the same way as how generalized
complex structures on a manifold are related with  Poisson
quasi-Nijenhuis structures. In the following two sections, we will
see that generalized complex structures on this Courant-Jacobi
algebroid unify the usual generalized complex structures on an
even-dimensional  manifold and generalized contact structures on an
odd-dimensional manifold.

\begin{defi}
A generalized complex structure on the Courant-Jacobi algebroid
$(A\oplus A^*,\pair{\cdot,\cdot},\jacobi{\cdot,\cdot},\rho)$ is a
bundle map $\huaJ:A\oplus A^*\longrightarrow  A\oplus A^*$
satisfying the algebraic properties
\begin{equation}\label{condition J}
\huaJ^2=-\Id,\quad\pair{\huaJ u,\huaJ v}=\pair{u,v},\quad\forall
~u,~v\in\Gamma(A)\oplus\Gamma(A^*)
\end{equation}
and the integrability condition
\begin{equation}\label{integ condition J}
\jacobi{\huaJ u,\huaJ v}-\jacobi{u,v}-\huaJ(\jacobi{\huaJ
u,v}+\jacobi{u,\huaJ v})=0,
\end{equation}
where $\pair{\cdot,\cdot}$ and $\jacobi{\cdot,\cdot}$ are given by
$(\ref{eqn:pair of Jacobi bi})$ and $(\ref{eqn:bracket theta 1})$
respectively.
\end{defi}
By (\ref{condition J}), $\huaJ$ must be of the form
\begin{equation}\label{J}
\huaJ=\Big(\begin{array}{cc}N&\pi^\sharp\\\sigma_\flat&-N^*\end{array}\Big),
\end{equation}
where $\pi\in\Gamma(\wedge^2A)$, $\sigma\in\Gamma(\wedge^2A^*)$,
$N\in\Gamma(A^*\otimes A)$, in which the following conditions are
satisfied:
\begin{eqnarray*}
N\circ\pi^\sharp=\pi^\sharp\circ N^*,\quad
N^2+\pi^\sharp\circ\sigma_\flat=-\Id,\quad
N^*\circ\sigma_\flat=\sigma_\flat\circ N.
\end{eqnarray*}

Similar as the proof of Proposition 2.2 in \cite{Marius}, we have
\begin{pro}
For any generalized complex structure $\huaJ$ given by (\ref{J}) on
the Courant-Jacobi algebroid $(A\oplus
A^*,\pair{\cdot,\cdot},\jacobi{\cdot,\cdot},\rho)$, $\pi$ is a
Jacobi bivector field. Thus there is an induced Jacobi structure on
the base manifold $M$.
\end{pro}
\begin{rmk}
The author gives his warmest thanks to the referee for pointing out
this fact.
\end{rmk}

We deform a Courant-Jacobi algebroid using a bundle map $\huaJ$.
More precisely, we introduce a new inner product
$\pair{\cdot,\cdot}_\huaJ$, a new bracket
$\jacobi{\cdot,\cdot}_\huaJ$ and a new anchor $\rho_\huaJ$ by
\begin{eqnarray*}
\pair{u,v}_\huaJ&=&\pair{\huaJ u,\huaJ v},\\
~\jacobi{u,v}_\huaJ&=&\jacobi{\huaJ u,v}+\jacobi{u,\huaJ v}-\huaJ\jacobi{u,v},\\
\rho_\huaJ&=&\rho\circ \huaJ.
\end{eqnarray*}

\begin{pro}\label{pro:J generalized}
Let  $\huaJ:A\oplus A^*\longrightarrow A\oplus A^*$ be a bundle
map given by (\ref{J}), then $\huaJ$ is a generalized complex
structure if and only if $(A\oplus
A^*,\pair{\cdot,\cdot}_\huaJ,[\cdot,\cdot]_\huaJ,\rho_\huaJ)$ is a
Courant-Jacobi algebroid such that $\huaJ$ is a Courant-Jacobi
algebroid morphism from $(A\oplus
A^*,\pair{\cdot,\cdot}_\huaJ,\jacobi{\cdot,\cdot}_\huaJ,\rho_\huaJ)$
to $(A\oplus A^*,\pair{\cdot,\cdot},\jacobi{\cdot,\cdot},\rho)$.
\end{pro}
\pf If $\huaJ$ given by (\ref{J}) is a generalized complex
structure, first we note that
$\pair{\cdot,\cdot}_\huaJ=\pair{\cdot,\cdot}$.
$\jacobi{\cdot,\cdot}_\huaJ$ is still a Leibniz bracket follows
from (\ref{integ condition J}). Also by (\ref{integ condition J}),
for any $u,v\in\Gamma(A\oplus A^*)$, we have
\begin{eqnarray*}
\rho_\huaJ(\jacobi{u,v}_\huaJ)=\rho\circ
\huaJ\jacobi{u,v}_\huaJ=\rho\jacobi{\huaJ u,\huaJ v}=[\rho\circ
\huaJ u,\rho\circ \huaJ v]=[\rho_\huaJ u,\rho_\huaJ v],
\end{eqnarray*}
 which implies $\rho_\huaJ$ is a homomorphism. Next we
verity that the conditions $(a),(b)$ in Definition
\ref{defi:CourantJacobi} are satisfied. Since $\huaJ$ preserves
the inner product $\pair{\cdot,\cdot}$, we have
\begin{eqnarray*}
\pair{\jacobi{u,v}_\huaJ,v}&=&\pair{\huaJ\jacobi{u,v}_\huaJ,\huaJ
v}=\pair{\jacobi{\huaJ u,\huaJ v},\huaJ v}=\pair{\huaJ
u,\jacobi{\huaJ v,\huaJ v}}=\pair{\huaJ
u,\huaJ\jacobi{v,v}_\huaJ}\\&=&\pair{u,\jacobi{v,v}_\huaJ},
\end{eqnarray*}
which implies that Condition $(a)$ in Definition
\ref{defi:CourantJacobi} is satisfied. Similarly, we have
$$
\rho_\huaJ(u)\pair{v,v}=\rho(\huaJ u)\pair{\huaJ v,\huaJ
v}=2\pair{\jacobi{\huaJ u,\huaJ v},\huaJ
v}=2\pair{\huaJ\jacobi{u,v}_\huaJ,\huaJ
v}=2\pair{\jacobi{u,v}_\huaJ,v},
$$
which implies that Condition $(b)$ is satisfied. Thus $(A\oplus
A^*,\pair{\cdot,\cdot}_\huaJ,\jacobi{\cdot,\cdot}_\huaJ,\rho_\huaJ)$
is a Courant-Jacobi algebroid. Furthermore, $\huaJ$ is a
Courant-Jacobi algebroid morphism from Courant-Jacobi algebroid
$(A\oplus
A^*,\pair{\cdot,\cdot}_\huaJ,\jacobi{\cdot,\cdot}_\huaJ,\rho_\huaJ)$
to $(A\oplus A^*,\pair{\cdot,\cdot},\jacobi{\cdot,\cdot},\rho)$ is
obvious. The converse part is straightforward and the proof is
completed. \qed

\begin{thm}\label{thm:Courant-Jacobi and Jacobi quasi}
Let $\huaJ:A\oplus A^*\longrightarrow A\oplus A^*$ be a bundle map
given by (\ref{J}). Then $(A\oplus
A^*,\pair{\cdot,\cdot}_\huaJ,[\cdot,\cdot]_\huaJ,\rho_\huaJ)$ is a
Courant-Jacobi algebroid if and only if $((A,\rho),\pi,N,\jetd
\sigma)$ is a Jacobi quasi-Nijenhuis algebroid.
\end{thm}
\pf One can easily see that for all $X,Y\in\Gamma(A)$ and
$\xi,\eta\in\Gamma(A^*)$, we have
\begin{eqnarray*}
[X,Y]_\huaJ&=&[X,Y]_N+\jetd\sigma(X,Y,\cdot),\\
~[\xi,\eta]_\huaJ&=&\Dorfman{\xi,\eta}_{\pi},\\
~[X,\xi]_\huaJ&=&[X,\pi^\sharp(\xi)]-\pi^\sharp\Lied_X\xi+\Lied_{NX}\xi-\Lied_X(N^*\xi)+N^*\Lied_X\xi,\\
~[\xi,X]_\huaJ&=&-[X,\xi]_\huaJ-\huaJ\jetd(\xi(X)).
\end{eqnarray*}
Therefore, if $(A\oplus
A^*,\pair{\cdot,\cdot}_\huaJ,[\cdot,\cdot]_\huaJ,\rho_\huaJ)$ is a
Courant-Jacobi algebroid, $A^*$ is a Dirac structure, and $A$ is its
isotropic transversal complement. By Theorem \ref{thm:quasi Liebi},
we obtain a quasi-Jacobi bialgebroid. More precisely, we have
\begin{eqnarray*}
\rho_A=\rho\circ N, \quad[\cdot,\cdot]_A=[\cdot,\cdot]_N, \quad
\delta=\jetd_N, \quad \phi=\jetd\sigma,
\end{eqnarray*}
and the quasi-Jacobi bialgebroid is given by
$((A^*,\rho\circ\pi^\sharp),\jetd_N,\jetd\sigma)$, or equivalently
$((A,\rho),\pi,N,\jetd\sigma)$ is a Jacobi quasi-Nijenhuis
algebroid.

Conversely, assume $((A,\rho),\pi,N,\jetd\sigma)$ is a Jacobi
quasi-Nijenhuis algebroid, then
$((A^*,\rho\circ\pi^\sharp),\jetd_N,\jetd\sigma)$ is a
quasi-Jacobi bialgebroid and its double is a Courant-Jacobi
algebroid, denote by $E$. It is straightforward to see that $E$ is
isomorphic to $(A\oplus
A^*,\pair{\cdot,\cdot}_\huaJ,\jacobi{\cdot,\cdot}_\huaJ,\rho_\huaJ)$.
\qed

By Proposition \ref{pro:J generalized} and Theorem
\ref{thm:Courant-Jacobi and Jacobi quasi}, we have
\begin{thm}
Let $(A,\rho)$ be a Jacobi algebroid. Assume that $\huaJ:A\oplus
A^*\longrightarrow A\oplus A^*$ is a bundle map given by
(\ref{J}), then $\huaJ$ is a generalized complex structure is
equivalent to that $((A,\rho),\pi,N,\jetd\sigma)$ is a Jacobi
quasi-Nijenhuis algebroid such that $\huaJ$ is a Courant-Jacobi
algebroid morphism from Courant-Jacobi algebroid $(A\oplus
A^*,\pair{\cdot,\cdot}_\huaJ,\jacobi{\cdot,\cdot}_\huaJ,\rho_\huaJ)$
to $(A\oplus A^*,\pair{\cdot,\cdot},\jacobi{\cdot,\cdot},\rho)$,
where the first one corresponds to the quasi-Jacobi bialgebroid
$((A^*,\rho\circ\pi^\sharp),\jetd_N,\jetd\sigma)$.
\end{thm}

\section{Generalized complex structures on $\huaT M$}
In this section, we consider the case where the vector bundle $A$
is the tangent bundle $TM$ of a manifold $M$. Since the tangent
Lie algebroid is a special Jacobi algebroid, it follows that
generalized complex structures on a manifold $M$ is a special case
of what we discussed in the last section. Next we first recall the
notion of generalized complex structures on a manifold $M$ and
then we deform the tangent Lie algebroid to be a Jacobi algebroid
and study its generalized complex structures. Consider the
generalized tangent bundle
$$\huaT M:=TM\oplus T^*M,$$ on its
section space $\Gamma(\huaT M)$, there is a well known Dorfman
bracket, explicitly,
\begin{equation}\label{eqn:bracket}
[X+\xi,Y+\eta]=[X,Y]+L_X\eta-L_Y\xi+\dM(\xi(Y)),\quad\forall~X+\xi,~Y+\eta\in\Gamma(\huaT).
\end{equation}
%Note that this bracket is not skew-symmetric but it does satisfy the Jacobi identity.
%Furthermore, there is a canonical non-degenerate symmetric bilinear form on $\huaT M$:
%\begin{equation}\label{eqn:pair}( X+\xi,Y+\eta)=\frac{1}{2}\big(\eta(X)+\xi(Y)\big)\end{equation}
%The notion of Courant algebroids was introduced in \cite{LWXmani}, $\huaT M$ with the bracket (\ref{eqn:bracket}) and the pairing
%(\ref{eqn:pair}) is the standard Courant algebroid associated with $M$.

\begin{defi}
A generalized complex structure on a manifold $M$ is a bundle map
$\huaJ:\huaT M\longrightarrow\huaT M$ satisfying the algebraic
properties:
$$\huaJ^2=-\Id\quad\mbox{and}\quad\langle\huaJ(u),\huaJ(v)\rangle=\langle u,v\rangle$$
and the integrability condition:
$$[\huaJ(u),\huaJ(v)]-[u,v]-\huaJ\big([\huaJ(u),v]+[u,\huaJ(v)]\big)=0,\quad\forall~u,~v\in\Gamma(\huaT).$$
%Here, $\langle\cdot,\cdot\rangle$ and $[\cdot,\cdot]$ are given by $(\ref{eqn:pair})$ and $(\ref{eqn:bracket})$ respectively.
\end{defi}

%Roughly speaking, generalized complex structures on $M$ of the form (\ref{J}) correspond to Poisson quasi-Nijenhuis manifolds,
%see \cite{Poisson quasi-Nijenhuis} for more details. It is a special case of generalized complex structures we study in last
%section where the Jacobi algebroid structure on $TM$ is trivial, i.e. the representation is just given by the anchor.  Next we consider more interesting cases.

We consider the bracket (\ref{eqn:bracket}) deformed by a
1-cocycle $\phi_0$ in the deRham cohomology. More precisely, the
new bracket $\jacobi{\cdot,\cdot}$ is given by
\begin{equation}\label{eqn:bracket theta}
\jacobi{X+\xi,Y+\eta}=[X,Y]+L_X\eta-L_Y\xi+\dM(\xi(Y))+(i_X\phi_0)\eta-i_Y(\phi_0\wedge\xi).
\end{equation}
It is easy to see that $(\Gamma(\huaT M),\jacobi{\cdot,\cdot})$ is
still a Leibniz algebra, but it is not a Courant algebroid since
$$
\jacobi{X+\xi,Y+\eta}=\dM(\xi(Y))+\xi(Y)\phi_0.
$$
In fact, $\phi_0$ decides a representation $\rho:TM\longrightarrow
TM\oplus \mathbb R$ which is given by
\begin{equation}\label{eqn:anchor jacobi on TM}
\rho(X)=X+\phi_0(X).
\end{equation}
Now $(TM,\rho)$ is a Jacobi algebroid. We  rewrite
(\ref{eqn:bracket theta}) as
\begin{equation}\label{eqn:bracket jacobi on TM}
\jacobi{X+\xi,Y+\eta}=[X,Y]+\Lied_X\eta-\Lied_Y\xi+\jetd(\xi(Y)).
\end{equation}
Therefore, we obtain a Courant-Jacobi algebroid $(\huaT
M,\pair{\cdot,\cdot},\jacobi{\cdot,\cdot},\rho)$, where
$\pair{\cdot,\cdot},~\jacobi{\cdot,~\cdot},\rho$ are given by
(\ref{eqn:pair of Jacobi bi}), (\ref{eqn:bracket jacobi on TM})
and (\ref{eqn:anchor jacobi on TM}) respectively.

\begin{pro}\label{pro:generalized on TM}
With the above notations, consider generalized complex structures
of  the Courant-Jacobi algebroid $(\huaT
M,\pair{\cdot,\cdot},\jacobi{\cdot,\cdot},\rho)$, we have
\begin{itemize}
\item[\rm(1).] For any $N:TM\longrightarrow TM$ which is a
Nijenhuis operator and satisfies $N^2=-\Id$,
$\Big(\begin{array}{cc}N&0\\0&-N^*\end{array}\Big) $is a
generalized complex structure.

\item[\rm(2).]For any $\omega\in\Omega^2(M)$,
$\Big(\begin{array}{cc}0&-\omega^{-1}\\\omega &0\end{array}\Big)$
is a generalized complex structure if and only if $\jetd\omega=0$.

\item[\rm(3).] For a $(1,1)$-tensor $N$ satisfying $N^2=-\Id$ and
$\pi\in\frkX^2(M)$,
$\Big(\begin{array}{cc}N&\pi\\0&-N^*\end{array}\Big)$ is a
generalized complex structure if and only if
\begin{eqnarray*}
\nonumber N\circ\pi^\sharp&=&\pi^\sharp\circ N^*,\\
\label{eqn:pi theta}[\pi^\sharp(\xi),\pi^\sharp(\eta)]&=&\pi^\sharp\Dorfman{\xi,\eta}_\pi,\\
\label{eqn:pi
a}N^*(\Dorfman{\xi,\eta}_\pi)&=&\Lied_{\pi^\sharp(\xi)}(N^*\eta)-\Lied_{\pi^\sharp(\eta)}(N^*\xi)-\jetd
\pi(N^*\xi,\eta),
\end{eqnarray*}
where $\Dorfman{\xi,\eta}_\pi$ is given by (\ref{eqn:bracket pi}).
\end{itemize}
\end{pro}

\begin{cor}If we write (2) and (3) in the above proposition in term of $\phi_0$, we have
\begin{itemize}
\item[\rm(1).] For any nondegenerate conformal symplectic
structure $(\phi_0,\omega)$, i.e. $\omega\in\Omega^2(M)$ is
nondegenerate and satisfies $\dM\omega=\phi_0\wedge\omega$,
$\Big(\begin{array}{cc}0&-\omega^{-1}\\\omega &0\end{array}\Big)$
is a generalized complex structure.

\item[\rm(2).]For a $(1,1)$-tensor $N$ satisfying $N^2=-\Id$ and
$\pi\in\frkX^2(M)$ satisfying
\begin{eqnarray*}
\nonumber N\circ\pi^\sharp&=&\pi^\sharp\circ N^*,\\
\label{eqn:pi theta}[\pi^\sharp(\xi),\pi^\sharp(\eta)]&=&\pi^\sharp[\xi,\eta]_\pi+\frac{1}{2}i_{\phi_0}(\pi\wedge\pi)(\xi,\eta)=0,\\
\label{eqn:pi
a}N^*([\xi,\eta]_\pi+\pi(\eta,\xi)\phi_0)&=&L_{\pi^\sharp(\xi)}(N^*\eta)-L_{\pi^\sharp(\eta)}(N^*\xi)-\dM
\pi(N^*\xi,\eta)+\pi(\eta,N^*\xi)\phi_0,
\end{eqnarray*}
$\Big(\begin{array}{cc}N&\pi\\0&-N^*\end{array}\Big)$ is a
generalized complex structure, where $[\xi,\eta]_\pi$ is given by
$$
[\xi,\eta]_\pi=L_{\pi^\sharp(\xi)}\eta-L_{\pi^\sharp(\eta)}\xi-\dM\pi(\xi,\eta).
$$
\end{itemize}
\end{cor}
\begin{rmk}
By $(1)$ in Proposition \ref{pro:generalized on TM}, we can see that
there are some generalized complex structures which are stable when
the bracket is deformed by $(\ref{eqn:bracket theta})$. By (2), we
see that how  a conformal symplectic structure on a manifold relates
with a generalized complex structure.
\end{rmk}

\section{Generalized complex structures on $\huaE^1(M)$}
Note that only even-dimensional manifolds can have generalized
complex structures. In \cite{Wade contact manifold and g}, the
authors give the odd-dimensional analogue of the concept of
generalized complex structures. Denote $(TM\oplus\mathbb
R)\oplus(T^*M\oplus\mathbb R)$ by $\huaE^1(M)$, and there is a
natural bilinear form $\langle\cdot,\cdot\rangle$ on $\huaE^1(M)$
defined by:
\begin{equation}\label{eqn:pair 2n+1}
\big\langle(X_1,f_1)+(\alpha_1,g_1),(X_2,f_2)+(\alpha_2,g_2)\big\rangle=\frac{1}{2}\big(\alpha_2(X_1)+\alpha_1(X_2)+f_1g_2+f_2g_1\big).
\end{equation}
There is also a bracket  which is given by
\begin{eqnarray}\label{eqn:bracket 2n+1}
&&\nonumber\jacobi{(X_1,f_1)+(\alpha_1,g_1),(X_2,f_2)+(\alpha_2,g_2)}\\
&=&([X_1,X_2],X_1f_2-X_2f_1)+
\widetilde{\Lied}_{(X_1,f_1)}(\alpha_2,g_2)-i_{(X_2,f_2)}\widetilde{d}(\alpha_1,g_1).
\end{eqnarray}
For more information about $\widetilde{\Lied}$ and
$\widetilde{d}$, see \cite{Wade contact manifold and g}.

\begin{defi}
A generalized contact structure on a (2n+1)-dimensional  manifold
$M$ is a bundle map $\huaJ:\huaE^1(M)\longrightarrow\huaE^1(M)$
satisfying the algebraic properties:
$$\huaJ^2=-\Id\quad\mbox{and}\quad\langle\huaJ(u),\huaJ(v)\rangle=\langle u,v\rangle$$
and the integrability condition:
$$\jacobi{\huaJ(u),\huaJ(v)}-\jacobi{u,v}-\huaJ\big(\jacobi{\huaJ(u),v}+\jacobi{u,\huaJ(v)}\big)=0,\quad\forall~u,~v\in\Gamma(\huaE^1(M)).$$
Here, $\langle\cdot,\cdot\rangle$ and $[\cdot,\cdot]$ are given by
$(\ref{eqn:pair 2n+1})$ and $(\ref{eqn:bracket 2n+1})$
respectively.
\end{defi}

We know that $ TM\oplus\mathbb R=\dev (M\times\mathbb R)$, the
covariant differential operator bundle of the trivial line bundle
$M\times\mathbb R$. In fact, we also have $T^*M\oplus\mathbb
R=~\jet (M\times\mathbb R)$, the first jet bundle of the trivial
line bundle $M\times\mathbb R$. In \cite{CLomni}, the authors
proved that for any vector bundle $E$, the first jet bundle $\jet
E$ may be considered as an $E$-dual bundle of $\dev E$, i.e.
\begin{eqnarray*}  {\jet E}  &\cong &
\set{\nu\in \Hom( \dev{E}  ,E )\,|\,  \nu(\Phi)=\Phi\circ
\nu(\Id_E),\quad\forall ~~ \Phi\in \gl(E )}  \subset \Hom( \dev{E}
,E ).
\end{eqnarray*}
We can introduce an $E$-valued pairing $\ppairingE{\cdot,\cdot}$
on $\dev E\oplus\jet E$ by
\begin{equation}\label{eqn:pair omni}
\ppairingE{\frkd+\mu,\frkt+\nu}=\frac{1}{2}\big(\mu(\frkt)+\nu(\frkd)\big)=\frac{1}{2}\big(\conpairing{\frkt,\mu}_E+\conpairing{\frkd,\nu}_E\big),\quad\forall~\frkd+\mu,~\frkt+\nu\in\dev
E\oplus\jet E.
\end{equation}
Furthermore,  for any
 $\frkd \in\Gamma(\dev E)$, the Lie derivative $\Lied_{\frkd}:\Gamma(\jet E)\longrightarrow\Gamma(\jet E)$
 is defined by:
\begin{eqnarray}\nonumber%\label{frkdmu}
\conpairing{\Lied_{\frkd}\mu,\frkd\pie}_{E}&\defbe&
\frkd\conpairing{\mu,\frkd\pie}_{E}-\conpairing{\mu,[\frkd,\frkd\pie]_{\dev}}_{E},
\quad\forall~ \mu \in \Gamma(\jet{E}), ~
~\frkd\pie\in\Gamma(\dev{E}).
\end{eqnarray}
On the section space $\Gamma(\dev E\oplus \jet E)$, we can define
a bracket as follows
\begin{eqnarray}\label{eqn:backet omni}
\jacobi{\frkd+\mu,\frkr+\nu}&\defbe&
[\frkd,\frkr]_{\dev}+\Lied_{\frkd}\nu-\Lied_{\frkr}\mu +
\jetd\mu(\frkr).
\end{eqnarray}
%\begin{defi}The quadruple $(\dev E\oplus \jet
%E,\jacobi{\cdot,\cdot},\ppairingE{\cdot,\cdot},p)$ is called an {\em omni-Lie algebroid},  where $p$ is the projection from $\dev
%E\oplus \jet E$ to $\dev{E}$, $\ppairingE{\cdot,\cdot}$ and $\jacobi{\cdot,\cdot}$ are given by $(\ref{eqn:pair omni})$ and
%$(\ref{eqn:backet omni})$ respectively. \end{defi}
%For more information about omni-Lie algebroids, see \cite{CLomni}, \cite{CLS1} and \cite{CLS2}. \vspace{3mm}

Therefore, we have $\huaE^1(M)=\dev (M\times\mathbb R)\oplus\jet
(M\times\mathbb R),$ and we can rewrite (\ref{eqn:bracket 2n+1})
by (\ref{eqn:backet omni}) and (\ref{eqn:pair 2n+1}) by
\begin{eqnarray}\label{eqn:pair 2n+1 new}
\langle\frkd+\mu,\frkt+\nu\rangle=\frac{1}{2}\big(\mu(\frkt)+\nu(\frkd)\big),\quad\forall~\frkd+\mu,~\frkt+\nu\in\dev
(M\times\mathbb R)\oplus\jet (M\times\mathbb R).
\end{eqnarray}
 The following proposition is straightforward.

\begin{pro}
The quadruple
$(\huaE^1(M),\langle\cdot,\cdot\rangle,\jacobi{\cdot,\cdot},\Id)$ is
a Courant-Jacobi algebroid, where $\langle\cdot,\cdot\rangle$ and
$[\cdot,\cdot]$ are given by (\ref{eqn:pair 2n+1 new}) and
(\ref{eqn:backet omni}) and $\Id(\frkd+\mu)=\frkd$. Therefore,
generalized contact structures on an odd dimensional manifold  is
exactly generalized complex structures on this Courant-Jacobi
algebroid.
\end{pro}

\begin{ex}\rm{
We consider generalized complex structures $\huaJ$  of the type
$\Big(\begin{array}{cc}N&0\\0&-N^*\end{array}\Big)$, where
$N:TM\oplus\mathbb R\longrightarrow TM\oplus\mathbb R$ is a bundle
map. Then the requirements are $N^2=-\Id$ and $T(N)=0$ which are
similar as the condition of a usual generalized complex structure.
More simply, if we consider
$N=\Big(\begin{array}{cc}\varphi&-Y\\\eta&0\end{array}\Big)$, where
$\varphi\in\Gamma(T^*M\otimes TM)$, $Y\in\frkX(M)$ is a vector field
and $\eta\in\Omega^1(M)$ is a $1$-form, then the condition
$N^2=-\Id$ is equivalent to
$$
\Big(\begin{array}{cc}\varphi^2-\eta\otimes Y& -\varphi(Y)\\
\eta\circ\varphi&-\eta(Y)\end{array}\Big)=-\Id.
$$
Therefore,
\begin{eqnarray}
\label{eqn:almost contact 1}\eta(Y)&=&\yi,\quad \varphi^2-\eta\otimes Y=-\Id,\\
\label{eqn:almost contact 2}\varphi(Y)&=&0,\quad
\eta\circ\varphi=0.
\end{eqnarray}
But, we should note that (\ref{eqn:almost contact 2}) follows from
(\ref{eqn:almost contact 1}). In fact, if $\eta(Y)=1$ and
\begin{equation}\label{eqn:almost contact 3}
\varphi^2(X)=-X+\eta(X)Y,\quad\forall~X\in \frkX(M),
\end{equation}
 first we have $ \varphi^2(Y)=0$. In (\ref{eqn:almost contact
 3}), substitute $X$ by $\varphi(Y)$, we obtain
 $\varphi(Y)=\eta(\varphi(Y))Y$. Acting by $\varphi$, we obtain
 $$
0=\varphi^2(Y)=\varphi(\eta(\varphi(Y))Y)=\eta(\varphi(Y))\varphi(Y)=\eta(\varphi(Y))^2Y,
 $$
which implies $\eta(\varphi(Y))=0$, and therefore $\varphi(Y)=0$.
Thus, $(\varphi,Y,\eta)$ is an {\bf almost contact structure}.
Furthermore, by straightforward computations, $T(N)=0$ is
equivalent to
$$T(\varphi)(X_1,X_2)+\dM
\eta(X_1,X_2)Y=0,\quad\forall~X_1,~X_2\in\frkX(M),$$ which is
equivalent to the condition that $(\varphi,Y,\eta)$ is a {\bf
normal contact structure}, where $T(\varphi)$ is the Nijenhuis
torsion of $\varphi$, see (\ref{eqn:Nijenhuis torsion}). }
\end{ex}
\begin{ex}\rm{
We consider generalized complex structures $\huaJ$  of the type
$\Big(\begin{array}{cc}0&\Upsilon\\\Theta&0\end{array}\Big)$, where
$\Theta:TM\oplus\mathbb R\longrightarrow T^*M\oplus\mathbb R$ and
$\Upsilon:T^*M\oplus\mathbb R\longrightarrow TM\oplus\mathbb R$ are
bundle maps. Evidently, $\huaJ^2=-\Id$ implies that
$\Upsilon=-\Theta^{-1}$. $\huaJ^*=-\huaJ$ implies $\Theta $ is
skew-symmetric. At last, from the integrability condition, we obtain
that $\jetd(\Theta)=0$. Since $\Theta$ is skew-symmetric,  we can
assume
$\Theta=\Big(\begin{array}{cc}\omega&\eta\\-\eta&0\end{array}\Big)$,
where $\omega\in\Omega^2(M)$ is a 2-form and $\eta\in\Omega^1(M)$ is
a 1-form such that $\eta\wedge\omega^n\neq0$ to insure that $\Theta$
is invertible.

If we let $\frac{\partial}{\partial t}$ as a basis of
$\Gamma(M\times\mathbb R)$ in $\Gamma(TM\oplus \mathbb R)$, then
any $\frkd\in\Gamma(TM\oplus \mathbb R)$ can be write as
$\frkd=X+f\frac{\partial}{\partial t}$ for some $X\in\frkX(M)$ and
$f\in C^\infty(M)$. Dually, any $\mu\in\Gamma(T^*M\oplus \mathbb
R)$ can be write as $\mu=\xi+g\dM t$. Then it is easy to see
$\Theta=\Big(\begin{array}{cc}\omega&\eta\\-\eta&0\end{array}\Big)\in\wedge^2\Gamma(T^*M\oplus
\mathbb R)$ is given by $\omega+\dM t\wedge\eta$. Since the
representation of the Jacobi algebroid $TM\oplus \mathbb R$ is the
identity map, we have $\jetd\yi=\dM t$. Thus we have
$$\jetd\Theta=\jetd(\omega+\dM t\wedge\eta)=\dM\omega+\dM t\wedge(\omega-\dM\eta).$$
So $\jetd\Theta=0$ precisely means that  $\omega-\dM \eta=0$, i.e.
$\omega=\dM \eta$. Since we also have $\eta\wedge\omega^n\neq0$,
it follows that $\eta$ is a {\bf contact structure}.}
\end{ex}

\end{document}